\documentclass[preprint]{aastex631}

\shorttitle{Solar wind turbulence around Mars}
\shortauthors{Romanelli et al.}
\usepackage{amsmath}

\usepackage{ulem}

\usepackage{multirow}

\begin{document}

\title{The Incompressible Magnetohydrodynamic Energy Cascade Rate Upstream of Mars: Effects of the Total Energy and the Cross-Helicity on Solar Wind Turbulence}

\correspondingauthor{Norberto Romanelli. Submitted to ApJ}
\email{norberto.romanelli@nasa.gov}
\author[0000-0001-9210-0284]{Norberto Romanelli}
\affiliation{Department of Astronomy, University of Maryland, College Park, MD, USA}
\affiliation{Planetary Magnetospheres Laboratory, NASA Goddard Space Flight Center, Greenbelt, MD, USA}
\author[0000-0002-1272-2778]{Nahuel Andr\'es}
\affiliation{CONICET - Universidad de Buenos Aires, Instituto de Física Interdisciplinaria y Aplicada (INFINA), Ciudad Universitaria, 1428 Buenos Aires, Argentina}
\affiliation{Universidad de Buenos Aires, Facultad de Ciencias Exactas y Naturales, Departamento de Física, Ciudad Universitaria, 1428 Buenos Aires, Argentina}
\author[0000-0002-2778-4998]{Gina A.  DiBraccio}
\affiliation{Planetary Magnetospheres Laboratory, NASA Goddard Space Flight Center, Greenbelt, MD, USA}
\author[0000-0003-1138-652X]{Jaye L. Verniero}
\affiliation{Heliospheric Physics Laboratory, NASA Goddard Space Flight Center, Greenbelt, MD, USA}
\author[0000-0002-1215-992X]{Jacob R. Gruesbeck}
\affiliation{Planetary Magnetospheres Laboratory, NASA Goddard Space Flight Center, Greenbelt, MD, USA}
\author[0000-0003-3255-9071]{Adam Szabo}
\affiliation{Heliospheric Physics Laboratory, NASA Goddard Space Flight Center, Greenbelt, MD, USA}
\author[0000-0002-6371-9683]{Jared R. Espley}
\affiliation{Planetary Magnetospheres Laboratory, NASA Goddard Space Flight Center, Greenbelt, MD, USA}
\author[0000-0001-5258-6128]{Jasper S. Halekas}
\affiliation{Department of Physics and Astronomy, University of Iowa, Iowa City, Iowa, USA}

\begin{abstract}
Solar wind turbulence is a dynamical phenomenon that evolves with heliocentric distance. Orbiting  Mars since September 2014, Mars Atmosphere and Volatile EvolutioN (MAVEN) offers a unique opportunity to explore some of its main properties beyond $\sim 1.38$ au. Here, we analyze solar wind turbulence upstream of Mars’s bow shock,  utilizing more than five years of magnetic field and plasma measurements. This analysis is based on two complementary methodologies: 1) the computation of  magnetohydrodynamic (MHD) invariants characterizing incompressible fluctuations; 2) the estimation of the  incompressible energy cascade rate at MHD scales (i.e., $\langle\varepsilon^{T}\rangle_{MHD}$). Our results show the solar wind incompressible fluctuations are primarily in a magnetically dominated regime, with the component travelling away from the Sun having a higher median pseudo-energy. Moreover, turbulent fluctuations have a total energy per mass of up to $\sim \,300$ km$^2$\,s$^{-2}$, a range  smaller than reported at 1 au. For these conditions, we determine the probability distribution function of $\langle\varepsilon^{T}\rangle_{MHD}$ ranges mainly between $\sim-1\times 10^{-16}$ and $\sim1\times 10^{-16}$ J$\,$m$^{-3}$ s$\,^{-1}$, with a median equal to $-1.8\times 10^{-18}$ J$\,$m$^{-3}$ s$\,^{-1}$, suggesting back-transfer of energy. Our results also suggest that $|\langle\varepsilon^{T}\rangle_{MHD}|$ is correlated with the total energy per mass of fluctuations and that the median of $\langle\varepsilon^{T}\rangle_{MHD}$ does not vary significantly with the cross-helicity. We find, however, that the medians of the inward and outward pseudo-energy cascade rates vary with the solar wind cross-helicity. Finally, we discuss these results and their  implications for future studies that can provide further insight into the factors affecting solar wind energy transfer rate. 
\end{abstract}

\section{Introduction}

Turbulence is a ubiquitous, multi-scale, and nonlinear phenomenon occurring in many space plasma environments \citep[e.g.,][]{M2011, Al2018, Po2018}. Solar wind turbulence has been investigated by means of analytical theoretical models, numerical simulations and analysis of spacecraft magnetic field and plasma observations at different heliocentric distances \citep[e.g.,][]{Br2013}. At the magnetohydrodynamics (MHD) scales, fully developed solar wind turbulence is partly characterized by the presence of an inertial range, where local energy transfer takes place throughout several spatial and temporal scales without dissipation. One method to quantify the effects of this phenomenon is to estimate the total energy cascade rate,  {$\varepsilon$}, i.e., the amount of energy per unit volume per unit time cascading across {different spatial} scales, resulting from nonlinear processes. In the classical hydrodynamic picture, this cascade of energy occurs from large to small scales, i.e., $\varepsilon>0$ \citep{K1941a}. {In plasma turbulence,} the energy cascade rate can be determined by means of exact relations, expressed in terms of plasma and magnetic field increment functions \citep[e.g.,][]{Br2013,M2023}. \citet{P1998a,P1998b} derived an exact relation valid for incompressible MHD fully developed turbulence, under statistical homogeneous and isotropic conditions. Numerical and observational results have evaluated and confirmed this relation \citep{Mi2009,SV2007,C2009b,St2009,St2011,He2018,H2018,Ba2020a,A2022}. Moreover, the inclusion of additional effects such as plasma compressibility, ion dynamics at smaller scales, and various thermodynamic closures enabled expressions for $\varepsilon$, valid for other plasma environments \citep{A2017b,A2018,A2019,A2019b,Ba2020b,A2021,M2022,B2023}.

Orbiting around Mars since September 2014, Mars Atmosphere and Volatile EvolutioN (MAVEN) offers an unique opportunity to explore the solar wind turbulent state at $\sim$ 1.5 au \citep[e.g.,][]{J2015,C2020, A2022}. \citet{Ru2017}  characterized the magnetic field power spectra upstream and inside the magnetosphere of Mars. The authors identified and characterized  variability in the power law index of magnetic fluctuations as a function of the magnetosphere region and Mars season. The latter aspect was mainly attributed to seasonal variability of the occurrence rate of waves observed at the local proton cyclotron frequency \citep[e.g.,][]{R2013,Ro2016,R2018c,Romeo2021}. \citet{Halekas2020_waves} computed the normalized solar wind cross-helicity and residual energy, utilizing MAVEN Magnetometer (MAG) and Solar Wind Ion Analyzer (SWIA) data \citep{Con2015,Ha2015b}. Consistent with previous studies, \citet{Halekas2020_waves} reported that Alfvénic fluctuations upstream of Mars are magnetically dominated, with higher pseudo-energy for the component traveling outwards from the Sun. \citet{A2020} estimated for the first time the absolute value of the incompressible energy cascade rate at MHD scales upstream of Mars ({i.e.,} $\langle|\varepsilon^{T}|\rangle_{MHD}$), using about four months of MAVEN observations. The authors observed changes in the probability distribution function (PDF) of $\langle|\varepsilon^{T}|\rangle_{MHD}$ with the Martian heliocentric distance and/or the presence of proton cyclotron waves (PCWs) \citep[see, e.g.,][]{R1990,B2002,M2004,R2013,Ro2016,Romeo2021,Romanelli2024}.  Analyzing more than five years of MAVEN data, \citet{Romanelli2022APJ} concluded that PCWs do not have a significant effect on $\langle|\varepsilon^{T}|\rangle_{MHD}$, and that the observed variability was due to changes in Mars’s heliocentric distance. 

To the best of our knowledge, so far there has not been a study computing the signed energy cascade rate at MHD scales upstream of Mars nor on its dependence on the total energy per mass of solar wind fluctuations and the solar wind cross-helicity. The main objective of the present paper is to improve the current understanding of the factors affecting the transfer of energy in the inertial range and to determine the PDF of the energy cascade rate beyond $\sim$1.38 au. These results are also put into context with previous reports in the inner heliosphere and provide added value, in particular, due to the region of the solar wind phase-space   nominally present upstream of Mars. 

This article is structured as follows: Section 2 presents a brief description of the incompressible MHD {equations utilizing} the Elsässer variables \citep{E1950} and also shows the exact relation {valid in the MHD inertial range} used to compute $\langle\varepsilon^{T}\rangle_{MHD}$. Section 3 reports the main capabilities of MAVEN MAG and SWIA instruments and the employed selection criteria to identify time intervals of interest and estimate $\langle\varepsilon^{T}\rangle_{MHD}$. Section 4 describes our main observational results and Section 5 develops our discussion and conclusions.

\section{Incompressible MHD Turbulence}\label{sec:model}

The incompressible MHD equations can be expressed in terms of the Elsässer variables $\mathbf{Z}^{\pm}$ \citep{E1950}, as follows:
\begin{equation}
	\frac{\partial \mathbf{{Z}^{\pm}}}{\partial t} = -(\mathbf{{Z}^{\mp}}\cdot\boldsymbol\nabla)\mathbf{{Z}^{\pm}} - \boldsymbol\nabla(P_{*}) + \nu \nabla^2\mathbf{{Z}^{\pm}},
 \label{Eq1n}
\end{equation}
{where $\mathbf{Z}^{\pm}=\mathbf{V}\pm \mathbf{V}_A$, $\mathbf{V}$ and $\mathbf{V}_A=\mathbf{B}/\sqrt{\mu_0 \rho_0}$ are the plasma flow and Alfvén velocity, respectively, $P_{*}$ is the total pressure and $\nabla \cdot \mathbf{{Z}^{\pm}}=0$}. Moreover, $\mathbf{B}$, $\rho_{0} = \langle \rho \rangle$ and $\mu_0$ are the magnetic field, the mean plasma mass density and the vacuum permeability, respectively. {The Elsässer variables $\mathbf{{Z}^{\pm}}$ describe general MHD processes, occurring for instance in the solar wind or planetary magnetosheaths \citep[e.g.,][]{S2020}. In particular, these terms can also be used to represent low-frequency waves, with  $\mathbf{{Z}^{-}}$ ($\mathbf{{Z}^{+}}$) associated with waves propagating parallel (antiparallel) to the background mean magnetic field.}

\subsection{Incompressible Solar Wind Fluctuations}

The incompressible solar wind fluctuations can be characterized by means of several parameters. In particular, the average total fluctuation energy per unit mass, $E_T$, can be expressed as: 
\begin{equation}
E_T=\frac{\langle |\Delta \mathbf{V}|^2\rangle+\langle|\Delta \mathbf{V}_A|^2\rangle}{2} = \frac{\langle|\Delta \mathbf{Z}^{+}|^2\rangle+\langle|\Delta \mathbf{Z}^{-}|^2\rangle}{4}.
 \label{Eq2n}
\end{equation}
where $\Delta \mathbf{V}$ and $\Delta \mathbf{V}_A$ are the mean solar wind (proton) and Alfvén velocity fluctuations, respectively. 
In other words, $E_T$ is the sum 
of the kinetic and magnetic fluctuation energies averaged over a given time interval, which is also equal to the sum of the pseudo-energies associated with Alfvénic fluctuations propagating anti-parallel ($\langle|\Delta \mathbf{Z}^{+}|^2\rangle$/4) and parallel ($\langle|\Delta \mathbf{Z}^{-}|^2\rangle$/4)) to the mean  magnetic field.

In addition, the normalized solar wind cross-helicity ($\sigma_c$) and residual energies ($\sigma_r$) are defined as follows,
\begin{equation}
\sigma_c=\frac{2\, \langle \Delta \mathbf{V} \cdot\Delta  \mathbf{V}_A \rangle}{\langle |\Delta \mathbf{V}|^2\rangle+\langle|\Delta \mathbf{V}_A|^2\rangle} = \frac{\langle|\Delta \mathbf{Z}^{+}|^2\rangle-\langle|\Delta \mathbf{Z}^{-}|^2\rangle}{\langle|\Delta \mathbf{Z}^{+}|^2\rangle+\langle|\Delta \mathbf{Z}^{-}|^2\rangle},
 \label{Eq3n}
\end{equation}
\begin{equation}
\sigma_r=\frac{\langle |\Delta \mathbf{V}|^2\rangle-\langle|\Delta \mathbf{V}_A|^2\rangle}{\langle |\Delta \mathbf{V}|^2\rangle+\langle|\Delta \mathbf{V}_A|^2\rangle},
 \label{Eq4n}
\end{equation}
In a few words, $\sigma_c$ quantifies the cross-correlation between the mean proton and Alfv\'en velocity fluctuations, or equivalently, the energy balance between anti-parallel and parallel propagating plasma fluctuations. $\sigma_r$ quantifies the energy balance between kinetic and magnetic field fluctuations. Moreover, by taking into account the interplanetary magnetic field (IMF) orientation for a given time interval, we can redefine the normalized solar wind cross-helicity to provide a measure of the energy balance between  fluctuations propagating inwards and outwards from the Sun \citep[see, e.g.,][]{Halekas2020_waves}.
In this study we compute $\sigma_c$ sign($\langle$IMF Bx$\rangle$), where Bx is the magnetic field component in the Mars Solar Orbital (MSO) coordinate system. Note that the MSO coordinate system is centered at Mars, with its x-axis pointing towards the Sun. The z-axis points northward and normal to the orbital plane, and the y-axis completes the right-handed coordinate system. Thus, the sign of the IMF Bx MSO component provides a very good estimation for the polarity of the radial IMF component. In this regard, the pseudo-energies associated with the components propagating inwards ({\it in}) and outwards ({\it out}) from the Sun,  $E_{in}$ and $E_{out}$, are equal to $\langle|\Delta \mathbf{Z}^{in}|^2\rangle$/4
 and $\langle|\Delta \mathbf{Z}^{out}|^2\rangle$/4, respectively \citep[see also, e.g.,][]{Stawarz_2010,Co2015,Vasquez_2018}.

{\subsection{Incompressible MHD Energy Cascade Rate}

As reported by \citet[][]{P1998a,P1998b}, the following exact relation for fully-developed incompressible three-dimensional MHD turbulence can be derived assuming homogeneous and isotropic conditions:}
\begin{equation}
   \langle  \delta \mathbf{Z}^{\mp}_{R}(L) \, |\delta \mathbf{Z}^{\pm}(L)|^2 \rangle \, {\rho_0}= -\frac{4}{3} \varepsilon^{\pm} \, L,
    \label{Eq5n}
\end{equation}
where $\delta \mathbf{Z}(L)= \mathbf{Z}(x+L)-\mathbf{Z}(x)$ refers to the difference of $\mathbf{Z}$ at two points separated by a distance $L$ along the radial ($\mathbf{R}$) direction, and $\mathbf{Z}^{\pm}_{R}=\mathbf{Z}^{\pm} \cdot \mathbf{R}$. Moreover, $\varepsilon^{\pm}$ corresponds to the energy cascade rates of the pseudo-energies  $|\mathbf{Z}^{\pm}|^2$, and the total nonlinear energy cascade rate is given by  $\varepsilon^{T} =(\varepsilon^{+}+\varepsilon^{-})/2$. The angular bracket $\langle\cdot\rangle$ refers to an ensemble average, which is computed as time average assuming ergodicity \citep{Po2001}. Under these definitions, positive and negative $\varepsilon^{T}$  values correspond to  direct and inverse energy transfer rates, respectively.

In the case of single-spacecraft measurements in the solar wind, all lagged separations are along the solar wind flow direction, $\mathbf{R}$.  Since we use the convention of positive differences, $\delta \mathbf{Z}^{\pm}(\tau) = - \delta \mathbf{Z}^{\pm}(L),$ where {$\tau$ is the timescale of interest,} {Eq.~\eqref{Eq5n}} can be rewritten as:
\begin{equation}
   \langle  \delta \mathbf{Z}^{\mp}_{R}(\tau) \, |\delta \mathbf{Z}^{\pm}(\tau)|^2 \rangle \, {\rho_0} = \frac{4}{3} \varepsilon^{\pm} \, V \, \tau
       \label{Eq6n}
\end{equation}
where we have made use of Taylor's hypothesis and the fact that the mean solar wind velocity ($V$) is much larger than typical velocity fluctuations. Same as before, by knowing the IMF orientation for each time interval under analysis, we can define  the pseudo-energy cascade rates inwards and outwards from the Sun at MHD scales ($\langle\varepsilon^{in}\rangle_{MHD}$ and $\langle\varepsilon^{out}\rangle_{MHD}$, respectively) as $\langle\varepsilon^{+}\rangle_{MHD}$ or $\langle\varepsilon^{+}\rangle_{MHD}$, accordingly. { Positive pseudo-energy cascade rates could be associated with a transfer of pseudo-energy from large to small scales within the solar wind. On the other hand, negative rates may indicate the presence of an inverse pseudo-energy cascade. Various mechanisms, such as large-scale shears, anisotropies, or a dominant wave mode, have been proposed to explain observed negative energy transfer rates in space plasmas and neutral fluids. Despite these efforts, further research is needed to better understand their nature. In particular, these analyses would benefit from a comprehensive characterization of small scale physical processes \citep[e.g.,][]{Co2015,Stawarz_2010,St2011,Verdini_2015,M2023,A2024}.}

\section{MAVEN MAG and SWIA Observations and selection criteria}
\label{sec:selcriteria}

To investigate the factors that affect the signed incompressible solar wind energy transfer rate at the MHD scales, we analyze MAVEN MAG and SWIA observations gathered between 10 October 2014 and 31 December 2019. MAVEN MAG measurements have a cadence of 32 Hz and an accuracy of $\sim$0.25 nT \citep{Con2015,MAG_data}. SWIA measures ion flux in the 25 eV/q to 25 keV/q energy range with a field of view of 360$^\circ{}$ by 90$^\circ{}$  \citep{Ha2015b,SWIA_data}. In this work, we have analyzed solar wind proton density and velocity observations, computed onboard with a sampling frequency of 0.25 Hz \citep{Halekas2017}.

The selection criteria and methodology employed in this work is analogous to the one used by \cite{Romanelli2022APJ}. What follows is a summary of its key steps. We initially identify $\sim$ 34 min intervals when MAVEN was upstream and magnetically disconnected from Mars' bow shock \citep{Gru2018,Greenstadt1986,Meziane2017}. { A sample of this size covers at least one correlation time of solar wind turbulent fluctuations \citep{Ma2018, H2017a}.} Next, we focus on cases with nearly incompressible solar wind conditions, i.e., $|n-n_0|/n_0<0.2$, where $n$ is the number plasma density and $n_0=\langle n \rangle$. For events satisfying this condition, the 32 Hz MAG data is averaged to derive values at times with available plasma moments (4s time resolution). This step is necessary to compute $\langle\varepsilon^{T}\rangle_{MHD}$ for every analyzed interval, as shown in Eq.~\eqref{Eq6n}. In addition, to determine a reliable energy cascade rate at MHD scales, $\langle\varepsilon^{T}\rangle_{MHD}$, we also restrict our analysis to events where the IMF cone angle (i.e., angle between the IMF and the solar wind velocity) is relatively stationary (variability equal or smaller than $15^\circ{}$), without changes of sign of the energy cascade rate throughout the MHD range and with $\frac{std(\varepsilon^{T}_{MHD})}{|\langle\varepsilon^{T}\rangle_{MHD}|}<1$ \citep[see, e.g.,][]{H2017a,A2020,A2021}. { To that end, we have also checked that the third-order moment displays a linear scaling with timescale, implying a nearly constant value of the energy cascade rates in the analyzed temporal scales.} Following \citet{A2020} and \citet{Romanelli2022APJ}, the computed values of  $\langle\varepsilon^{T}\rangle_{MHD}$ shown in the next sections are  the average of $\varepsilon^{T}$ between $\tau=500$ s and $\tau=1500$ s.  However, we report that no significant differences are present when computing $\langle\varepsilon^{T}\rangle_{MHD}$ based on averages between $\tau=750$ s and $\tau=1250$ s, {suggesting consistency across the inertial range \citep[see, e.g.,][]{Co2015}}. { We have also investigated the convergence of higher-order moments by assessing the variability of energy transfer rate estimates utilizing time intervals of different sizes. In this regard, it is important to note the constraints imposed by MAVEN's orbital period, the duration the spacecraft spends in the pristine solar wind without magnetic connection to the bow shock, and the requirement for a statistically significant number of events. These factors allow us to estimate the maximum size of the time intervals for computing the energy cascade rate, based on the current MAVEN data set. Taking these constraints into account, we analyze MAVEN observations based on the identification of 1890 intervals of $\sim$34 min and 376 intervals of $\sim$68 min satisfying the selection criteria. An analysis of the 
statistical outcomes obtained from these two data subsets does not show significant differences in the observed trends and associated conclusions. Hereafter, we therefore display results derived from all selected $\sim$34 min intervals. }

\section{Results}\label{sec:results}

\subsection{ Case Study: Turbulent Event on July 15, 2015}

\begin{figure}
\begin{center}
\includegraphics[scale=.70]{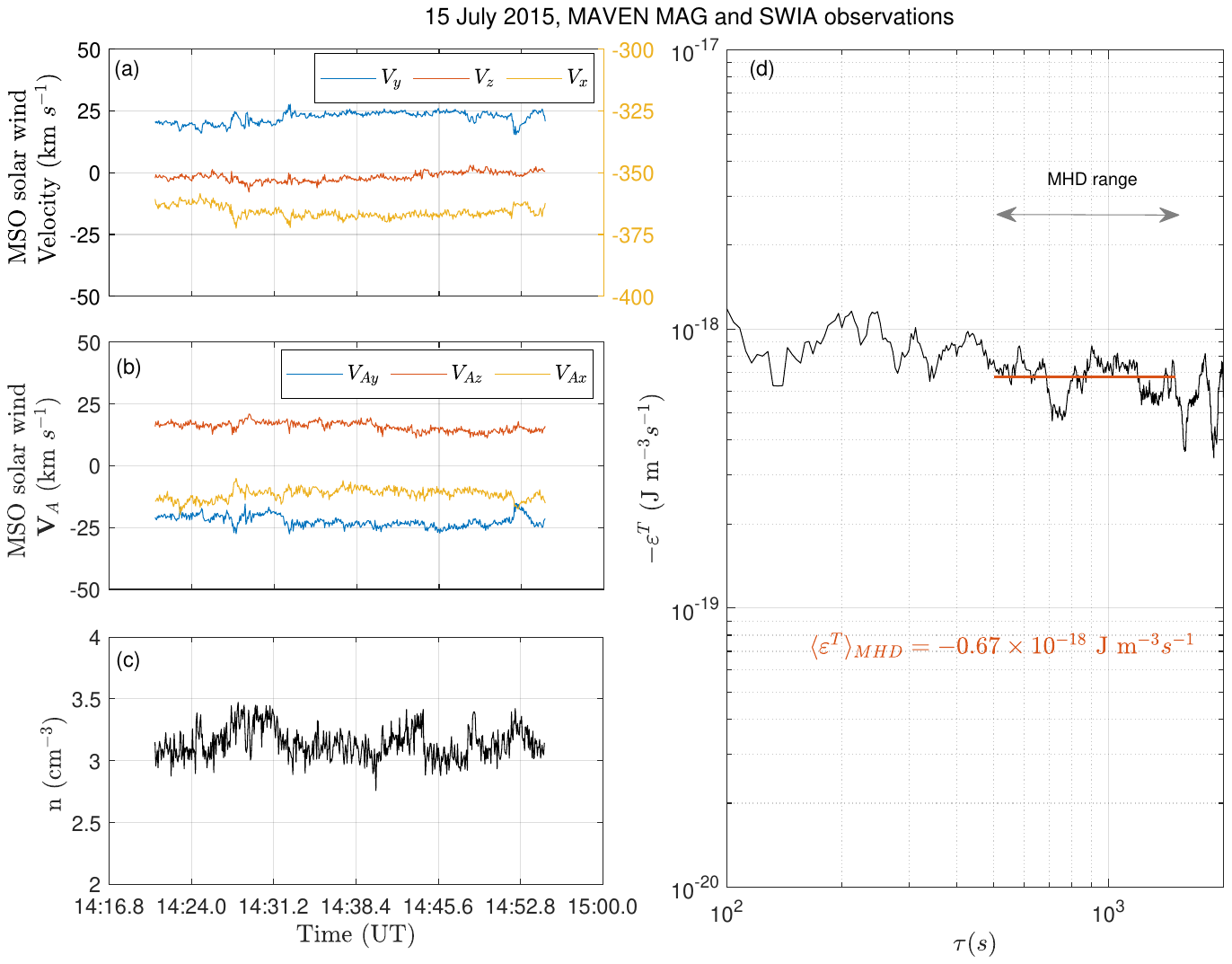}
\end{center}
\caption{Time series for an event observed by MAVEN in the pristine solar wind in 2015 July 15 (a): solar wind velocity components in the MSO coordinate system (left axis: $V_y$, $V_z$; right axis: $V_x$); (b) solar wind Alfv\'en velocity components in the MSO coordinate system; (c) solar wind number proton density; (d) total energy cascade rate $\varepsilon^{T}$ times (-1), as a function of the time lag $\tau$. The value of $\langle\varepsilon^{T}\rangle_{MHD}$ is derived from the average of $\varepsilon$ for  $\tau$ between 5$\times10^2$ s and 1.5$\times10^3$ s (as indicated by the horizontal double arrow).}
\label{fig1}
\end{figure}

{ An example of an event fulfilling our selection criteria  described in Section \ref{sec:selcriteria} is shown in Figure \ref{fig1}. It displays MAVEN MAG and SWIA observations obtained on July 15, 2015, from $\sim14:21$ UT to $\sim14:55$ UT. The solar wind velocity (panel a) is mostly anti-parallel to the x-MSO axis, and displays  an approximately constant value  throughout this interval with $V_x\sim-366$ km s$^{-1}$. Panel (b) indicates that the Alfvén velocity also remains relatively steady. Panel (c) shows the event took place under nearly incompressible conditions, with $n\sim$ 3.1 cm$^{-3}$. Panel (d) displays the total signed incompressible energy cascade rate as a function of $\tau$, with $\tau$ ranging from $10^2$ s to $2\times10^3$ s. We estimate $\langle\varepsilon^{T}\rangle_{MHD}$ is approximately $-0.67\times10^{-18}$ J m$^{-3}$ s$^{-1}$, suggesting  the presence of back-transfer of energy  in this event.}

\subsection{Solar Wind Incompressible Fluctuations}

\begin{figure}
\begin{center}
\includegraphics[scale=.85]{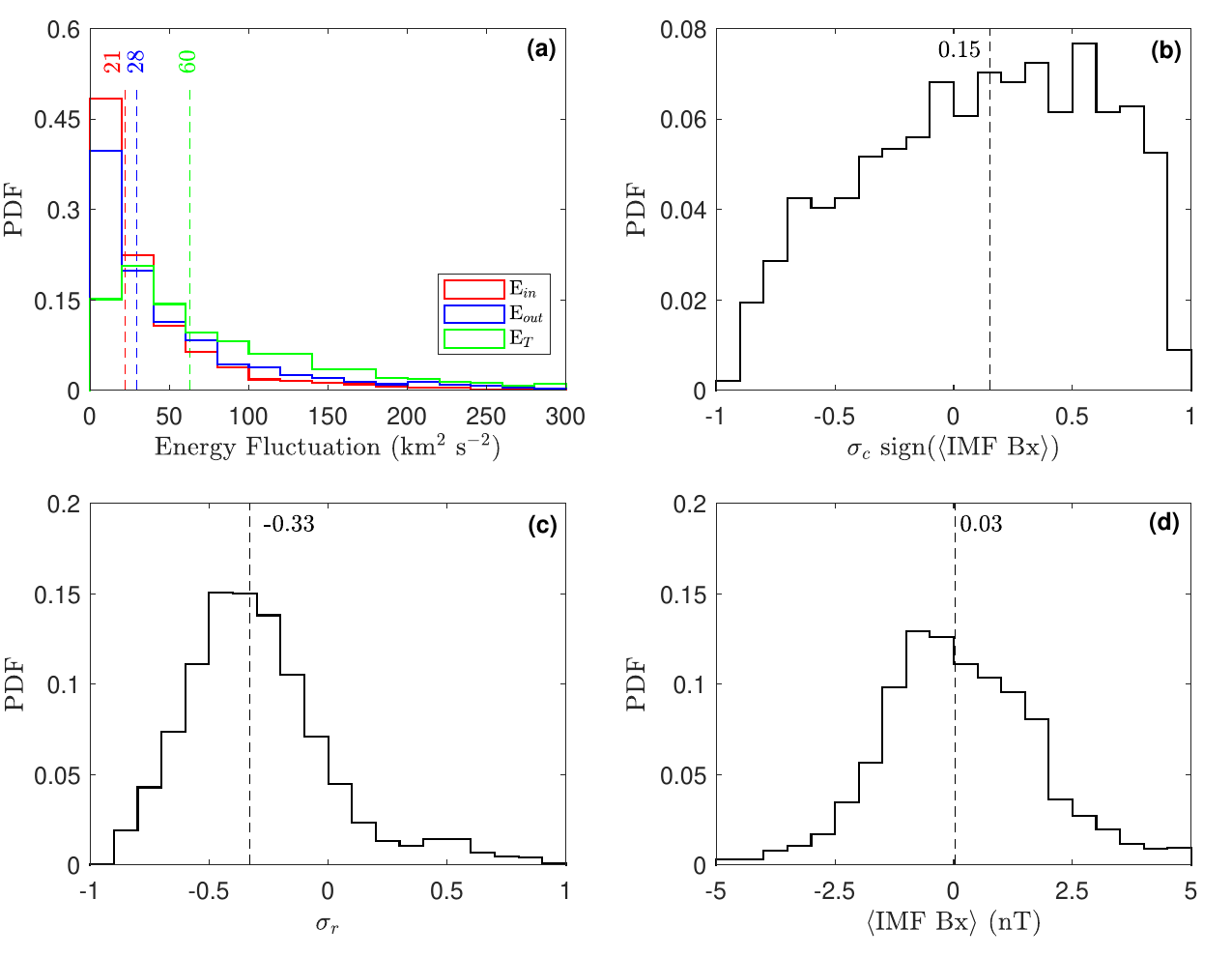}
\end{center}
\caption{Probability distribution functions for (a) pseudo-energies and total energy per mass of solar wind turbulent fluctuations ($E_{in}$, $E_{out}$ and $E_{T}$), (b) the cross-helicity ($\sigma_c$) times the sign of the IMF $B_x$ Mars Solar Orbital (MSO) component, (c) the residual energy ($\sigma_r$), (d) and the IMF $B_x$ MSO component. The vertical dashed lines in each panel correspond to the median of each distribution.}
\label{fig2}
\end{figure}

Figure \ref{fig2} displays the  PDF of (a) the total energy per mass of solar wind fluctuations and the pseudo-energies $E_{in}$ and $E_{out}$, (b) the normalized solar wind cross-helicity $\sigma_c$ times the sign($\langle$IMF Bx$\rangle$), (c) the residual energy and (d) the $\langle$IMF Bx$\rangle$ MSO component for all analyzed events observed by MAVEN. { These variables are computed based on averages over the size of the interval, i.e., $\sim 34$ min.}
Note that hereafter we refer to the total energy per mass as total energy. Our results show that the total energy of the turbulent  fluctuations extends up to 300 km$^2$s$^{-2}$, with a median of $\sim 60$ km$^2$s$^{-2}$ (green dashed line, Figure \ref{fig2} (a)). These values are significantly smaller than their reported counterparts at Earth or smaller heliocentric distances. This allows us to investigate solar wind turbulence in a scarcely explored region of  phase-space \citep[e.g.,][]{St2009, St2011,A2022}. The median pseudo-energy of the outward fluctuation component, $E_{out}$, is $\sim$ 28 km$^2$s$^{-2}$, $\sim 33\%$ larger than that of $E_{in}$ ($\sim$ 21 km$^2$s$^{-2}$). As a result, $\sigma_c$ sign($\langle$IMF Bx$\rangle$) displays a positive median ($\sim 0.15$), suggesting the solar wind turbulence state can be understood in terms of a combination of fluctuations travelling towards and away from the Sun, where the latter component is more energetic. Figure \ref{fig2} (c) shows most of the events are magnetically dominated (i.e., magnetic fluctuation energy larger than kinetic fluctuation energy), with a median of $\sigma_r \sim -0.33$. Figure \ref{fig2} (d) shows a nearly symmetric distribution (skewness $\sim 0.18$), with a median equal to 0.03 nT, suggesting MAVEN visited both sides of the heliospheric current sheet in a similar proportion and there are no significant  biases associated.  The quartiles ($Q_1$, $Q_2$, and $Q_3$) corresponding  to 
the distributions shown in Figure \ref{fig2} are provided in Table 1 {(see columns associated with $\sim 34$ min intervals).}

\subsection{The Solar Wind Incompressible MHD Energy Cascade Rate at Mars}

Figure \ref{fig3} shows the PDFs of $\langle\varepsilon^{in}\rangle_{MHD}$, $\langle\varepsilon^{out}\rangle_{MHD}$ and $\langle\varepsilon^{T}\rangle_{MHD}$ between $-1\times10^{-16}$ and $1\times 10^{-16}$ J$\,$m$^{-3}$ s$\,^{-1}$. The vertical lines correspond to the quartiles of each distribution. Our results show that the three distributions take a wide range of values and have negative medians, within the $95\%$ confidence interval. Specifically, the median for $\langle\varepsilon^{T}\rangle_{MHD}$ upstream of Mars is $-1.8\times 10^{-18}$ J$\,$m$^{-3}$ s$\,^{-1}$, suggesting the presence of back-transfer of energy at $\sim$ 1.5 au. { Interestingly, this conclusion holds when increasing the size of the analyzed time intervals. Indeed, we also observe a negative median for the MHD energy transfer rate distribution when the time interval duration is doubled. In other words, we obtained similar statistical results from an analogous analysis of MAVEN observations based on 376 intervals of $\sim$68 min. Moreover, the quartiles associated with the distribution of the pseudo-energy components and total energy cascade rates do not vary significantly with both interval sizes, as shown in Table 1.} In addition, it is worth mentioning that negative transfer rates have been found in studies focused on solar wind turbulence at 1 au, but only under certain solar wind conditions \citep[see,][]{Smith2009,Stawarz_2010,H2017a}. 

\subsubsection{Dependence on the Total Energy of Solar Wind Fluctuations}

Figure \ref{fig4} (a) displays a scatter plot of all analyzed events with $\sigma_r$ as a function of $\sigma_c$ sign($\langle$IMF Bx$\rangle$), and color-coded with the total energy of solar wind fluctuations. As also shown in Figure \ref{fig2}, we observe that most of the events are magnetically dominated, and with a slightly larger proportion of events with more pseudo-energy in the component going out of the Sun. In addition, we identify a weak trend in which the events with larger total fluctuation energy appear mostly distributed near the boundary ($\sigma_c^2+\sigma_r^2 \sim 1$). This can be understood in terms of the phase between the solar wind velocity and magnetic field  fluctuations. Indeed, these events present a relatively high correlation between these two fields \citep[see Eq.~7 in][]{Ba1998}.

Figure \ref{fig4} (b) shows the total solar wind energy cascade rate at the MHD scales as a function of the total energy of the turbulent fluctuations, in log-log scales. Our results display a strong positive correlation (R$=0.80$) between $\langle\varepsilon^T\rangle_{MHD}$ and $E_{T}$, when $\langle\varepsilon^T\rangle_{MHD}>0$ (grey dots). An analogous dependence is identified between $|\langle\varepsilon^T\rangle_{MHD}|$ and $E_{T}$, when $\langle\varepsilon^T\rangle_{MHD}<0$ (orange dots). Moreover, the computed linear fits suggest the presence of a polynomial dependence, with $|\langle\varepsilon^T\rangle_{MHD}|$ $\propto E_{T}^\alpha$, where $\alpha=1.3\pm0.1$ and $\alpha=1.2\pm0.1$, for each data set, respectively. The dispersion observed in both data sets is likely associated with other plasma variables affecting the solar wind turbulent cascade rate, such as, the solar wind cross-helicity.

\subsubsection{Dependence on the Cross-Helicity of Solar Wind Fluctuations}

Figure \ref{fig5} shows the medians of the energy cascade rates (a-c), the $\langle\varepsilon^{out}\rangle_{MHD}/\langle\varepsilon^{in}\rangle_{MHD}$ ratio (d-f), and the total and pseudo-energy of fluctuations (g-i) as a function of $\sigma_c$ sign($\langle$IMF Bx$\rangle$), { with each column corresponding to a different total energy fluctuation ($E_{T}$) range. As can be seen in panel (g) (which includes all analyzed events), the medians of $E_{out}$ (blue curve), $E_{in}$ (red curve), and $E_{T}$ (green curve) exhibit strong variation with the solar wind cross-helicity. Specifically, the median of $E_{T}$ is higher for highly Alfv\'enic ($|\sigma_c|\sim1$) solar wind fluctuation states. Also, as expected from Eq.~\eqref{Eq3n}, the median of $E_{in}$ ($E_{out}$) decreases (increases) with $\sigma_c$ sign($\langle$IMF Bx$\rangle$). It is worth emphasizing that these trends affect the observed variability of the solar wind energy cascade rates as a function of cross-helicity, as suggested by Figure \ref{fig4} (b).}

{ 
Motivated by these results, we further analyze two data subsets defined in terms of narrower ranges of total energy $E_{T}$. By doing this, we can examine if there is a dependence of the energy cascade rate with the cross-helicity. The central column (i.e., Figure \ref{fig5} (b, e, h)) considers events with $E_{T}\leq Q_3(E_{T}) \sim$ 120 km$^2\,$s$^{-2}$, while the right column (i.e., Figure \ref{fig5} (c, f, i)) focuses on events with $E_{T}\geq Q_1(E_{T})\sim$ 30 km$^2\,$s$^{-2}$. Our results suggest that the total energy of solar wind fluctuations (green curve) varies significantly less with solar wind cross-helicity for these two energy ranges (Figure \ref{fig5} (h, i)). In addition, we find mostly negative and approximately constant median total energy cascade rates at MHD scales across cross-helicity bins (Figure \ref{fig5} (b, c)).  Furthermore, the medians of $\langle\varepsilon^{in}\rangle_{MHD}$ and $\langle\varepsilon^{out}\rangle_{MHD}$ are positively and negatively correlated with solar wind cross-helicity, respectively, across all explored energy ranges (Figure \ref{fig5} (a-c)). As a result, the median for $\langle\varepsilon^{out}\rangle_{MHD}/\langle\varepsilon^{in}\rangle_{MHD}$ takes small positive values only for nearly non-correlated solar wind velocity and magnetic field fluctuations. These profiles are highly asymmetric, with negative (but small) values for $\sigma_c$ sign($\langle$IMF Bx$\rangle)<0$ and highly negative values for $\sigma_c$ sign($\langle$IMF Bx$\rangle)>0$ (Figure \ref{fig5} (d-f)).}

\section{Discussion and Conclusions}\label{sec:conclusions}

Making use of the incompressible MHD exact law and MAVEN magnetic field and plasma observations, we investigated the properties of solar wind Alfv\'enic fluctuations and determined the solar wind energy cascade rate at the MHD scales, upstream of the Martian bow shock. Solar wind turbulence properties at the Martian heliocentric distances, i.e., $\sim 1.38 -1.67$ au, display both similarities and differences with previous reports upstream of Earth's magnetosphere and closer to the Sun. 

\begin{figure}
\begin{center}
\hspace{-1cm}
\includegraphics[scale=.65]{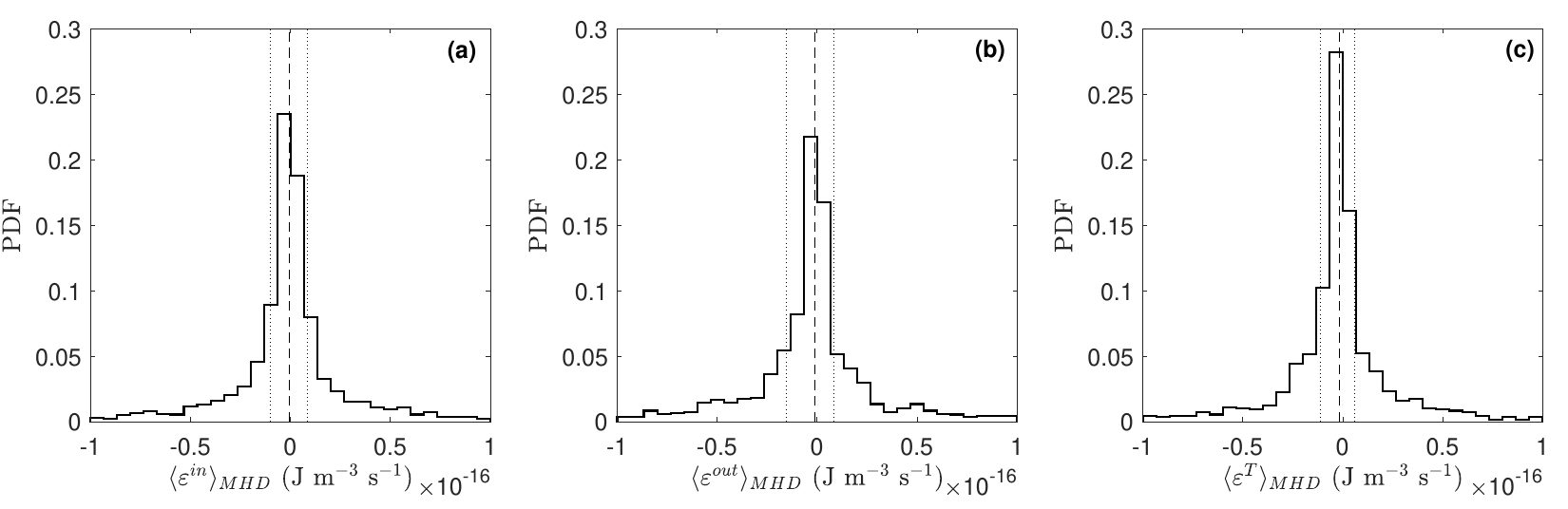}
\end{center}
\caption{Probability distribution functions for (a-c) $\langle\varepsilon^{in}\rangle_{MHD}$, $\langle\varepsilon^{out}\rangle_{MHD}$, and $\langle\varepsilon^{T}\rangle_{MHD}$, respectively. Vertical dashed (dotted) lines correspond to the respective medians (lower and upper quartiles) of each distribution. }
\label{fig3}
\end{figure}

Among the similarities, we find that most of the  analyzed events are characterized by negative residual energies, with a positive median for the solar wind cross-helicity \citep[e.g.,][]{Ba1998,B2005,C2020,Al2022,Halekas2020_waves,A2022}. On the other hand, the total  energy of solar wind fluctuations takes values significantly smaller than observed upstream of Earth's bow shock and at smaller heliocentric distances \citep[e.g.,][]{B2005,Stawarz_2010,Vasquez_2018}. We also find that the PDFs of $\langle\varepsilon^{in}\rangle_{MHD}$, $\langle\varepsilon^{out}\rangle_{MHD}$, and $\langle\varepsilon^{T}\rangle_{MHD}$ range mainly between $\sim-1\times 10^{-16}$ and $\sim1\times 10^{-16}$ J$\,$m$^{-3}$ s$\,^{-1}$ and have negative medians, suggesting transfer of energy from the smallest to the largest scales of the system in the studied temporal scales for slightly more than half of the analyzed events (see Figure \ref{fig3}). These results appear to be in contrast with some previous studies focused on solar wind turbulence upstream of Earth's bow shock, where a positive energy cascade rate is typically observed \citep[e.g.,][]{So2007,M2008,St2009}. However, negative solar wind transfer rates at MHD scales were observed at 1 au, on average, under certain solar wind conditions \citep[e.g.,][]{H2017a,Smith2009,Stawarz_2010}. Moreover, a partial explanation of our results can be provided based on the reports by \citet{Smith2009} and \citet{Stawarz_2010}. By applying the exact relation for incompressible MHD turbulence, \citet{Smith2009} and  \citet{Stawarz_2010} reported a significant back-transfer of solar wind energy from small to large scales for events with large absolute values of cross-helicity, upstream of the terrestrial bow shock. Interestingly, \citet{Stawarz_2010} also reported that the range of solar wind cross-helicity values where negative energy cascade rates are observed increases in size with decreasing total energy of turbulent fluctuations (see Figure 3 in \citet{Stawarz_2010}). In this regard, the relatively low total energy of solar wind fluctuations usually seen upstream of Mars provides an explanation for the negative median energy transfer rates observed at MHD scales. Indeed, the lowest total energy level analyzed by \citet{Stawarz_2010} is four times larger than the maximum value investigated here. As a result, our analysis shows negative median energy cascade rates for the entire cross-helicity range. 

Previous studies have characterized the evolution of the solar wind normalized cross-helicity, residual energy and energy of fluctuations as a function of heliocentric distance \citep[e.g.,][]{B2005,C2020,Al2022}. Overall, the solar wind cross-helicity displays a decreasing trend with relatively highly values ($\sim 0.6$) for small heliocentric distances ($\sim 0.2$ au), and appears to reach a small asymptotic positive value at  approximately 1 au. In other words, the non-linear interaction between solar wind fluctuations is responsible for the evolution from a highly Alfvenic state towards another one with small correlation between solar wind velocity and magnetic field fluctuations. This evolution takes places with magnetic field fluctuations dominating over velocity fluctuations, i.e., a relatively slowly varying but negative residual energy ($\sim -0.3$). In addition, the energy of fluctuations also decreases with heliocentric distance, with ${(\delta \mathbf{Z}^{+})}^2 > {(\delta \mathbf{Z}^{-})}^2$. Furthermore, the dependence of the pseudo-energy of these two components varies differently with heliocentric distance \citep[e.g.,][]{B2005,C2020}. Our analysis of MAVEN observations suggest these solar wind properties appear to evolve similarly with heliocentric distance beyond 1 au and, at least, up to Mars's orbital location. 
Observational studies have also shown that the intensity of the energy transfer rate at MHD scales increases
as the heliocentric distance decreases \citep[see, e.g.,][]{Ba2020b,A2021,A2022,Romanelli2022APJ,B2023}. In particular,  \citet{Ba2020b} reported that the energy transfer rate around the first Parker Solar Probe (PSP) perihelion is approximately 100 times the typical value at 1 au. \cite{A2022} analyzed more than two years of PSP observations and found that the absolute value of incompressible energy cascade rate is negatively correlated with the heliocentric distance. 
Similarly, \citet{B2023} observed an increase in the incompressible and compressible energy cascade rates as PSP approached the Sun, which they attributed to an increase in the solar wind fluctuations total energy.
The energy cascade rates observed upstream of Mars in the present work are consistent with this trend and  previous results obtained by \citet{Romanelli2022APJ}.

We also investigated the influence the total energy of solar wind turbulent fluctuations and the cross-helicity have on the total energy cascade rate and its components, $\langle\varepsilon^{in}\rangle_{MHD}$ and $\langle\varepsilon^{out}\rangle_{MHD}$. Our observational results suggest the median of the total energy cascade rate is not significantly affected by the solar wind cross-helicity at the Martian heliocentric distances, for relatively narrow total energy bins (see, Figure \ref{fig5} (b-c)). Indeed, most of the observed variability is associated with the dependence of $E_{T}$ on $\sigma_c$ sign($\langle$IMF Bx$\rangle$) (Figure \ref{fig5} (g)).  In contrast, we also find that the medians of  $\langle\varepsilon^{in}\rangle_{MHD}$, $\langle\varepsilon^{out}\rangle_{MHD}$, and $\langle\varepsilon^{out}\rangle_{MHD}/\langle\varepsilon^{in}\rangle_{MHD}$ vary with the solar wind cross-helicity. 
In particular, highly Alfv\'enic states are observed under relatively intense (negative) pseudo-energy cascade rates, $\langle\varepsilon^{in}\rangle_{MHD}$ and $\langle\varepsilon^{out}\rangle_{MHD}$ \citep[see also, e.g.,][]{Smith2009,Stawarz_2010}. Moreover, the ratio $\langle\varepsilon^{out}\rangle_{MHD}/\langle\varepsilon^{in}\rangle_{MHD}$ displays a non-symmetric profile with respect to its maximum, taking positive values only when  $\sigma_c$ sign($\langle$IMF Bx$ \rangle)$ is small, at least for the energy of fluctuations typically seen upstream of Mars's magnetosphere.
This analysis provides added value to previous reports focused on solar wind turbulent conditions upstream of Earth \citep[e.g.,][]{Co2015,Stawarz_2010}. In particular, the observed trends in Figure \ref{fig5} (d-f) are similar to what has been reported by \citet{Stawarz_2010} and \citet{Co2015}. However, a detailed comparison is limited, since these authors analyzed the relationship between the pseudo-energy cascade rates and the absolute value of the solar wind cross-helicity and/or   considered other cross-helicity ranges.

\begin{figure}
\begin{center}
\includegraphics[scale=.65]{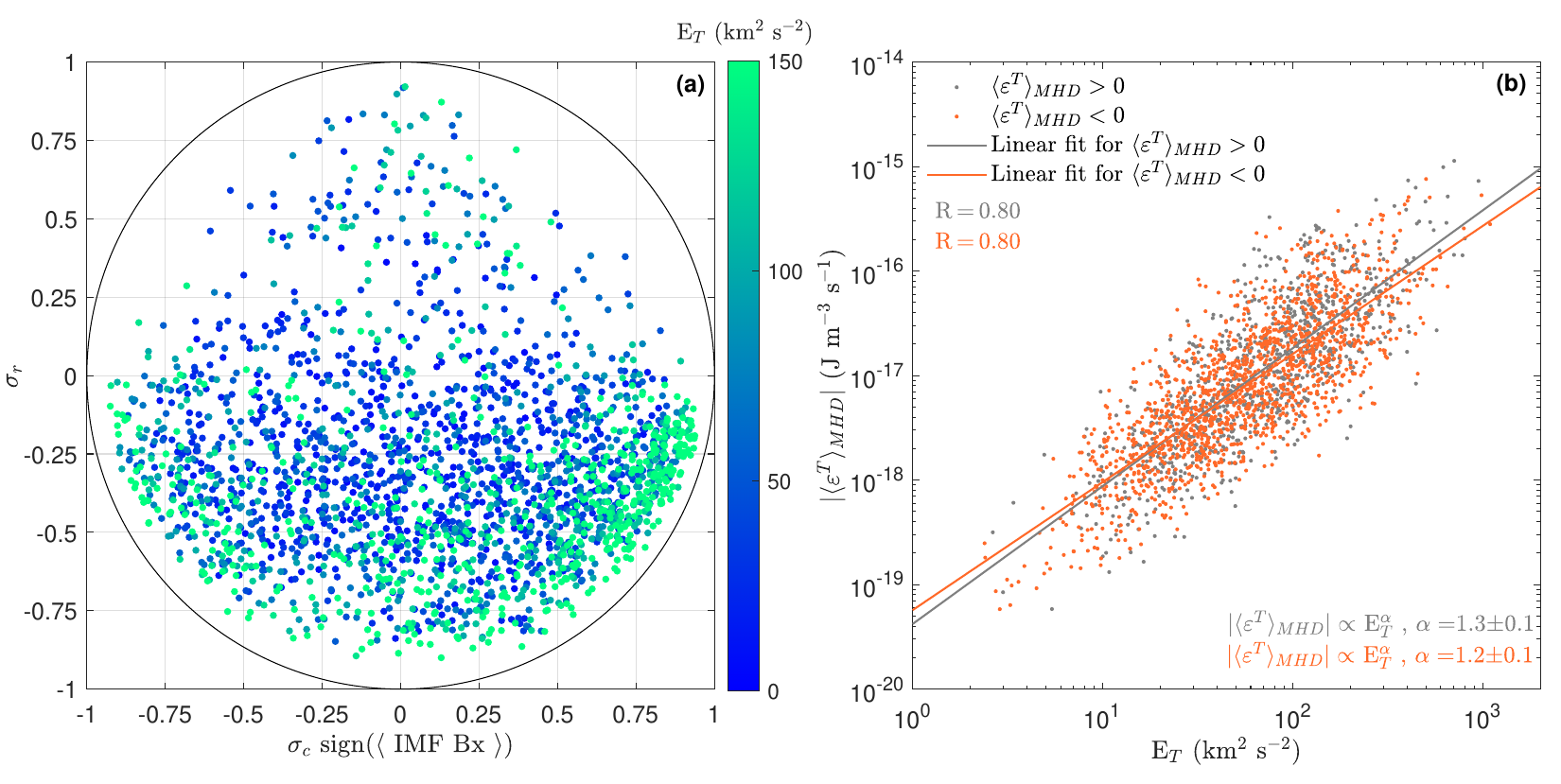}
\end{center}
\caption{(a) $\sigma_r$ as a function of $\sigma_c$ sign($\langle$IMF Bx$\rangle$), color-coded according to the total energy of solar wind turbulent fluctuations ($E_{T}$). (b) Total energy cascade rate at MHD scales ($\langle\varepsilon^T\rangle_{MHD}$) as a function of $E_{T}$. Events with positive  $\langle\varepsilon^T\rangle_{MHD}$ are shown in grey. Events with negative $\langle\varepsilon^T\rangle_{MHD}$ (orange points) are multiplied by a factor $(-1)$ to displayed them in a log-log scales.}
\label{fig4}
\end{figure}

\begin{figure}
\begin{center}
\includegraphics[scale=.85]{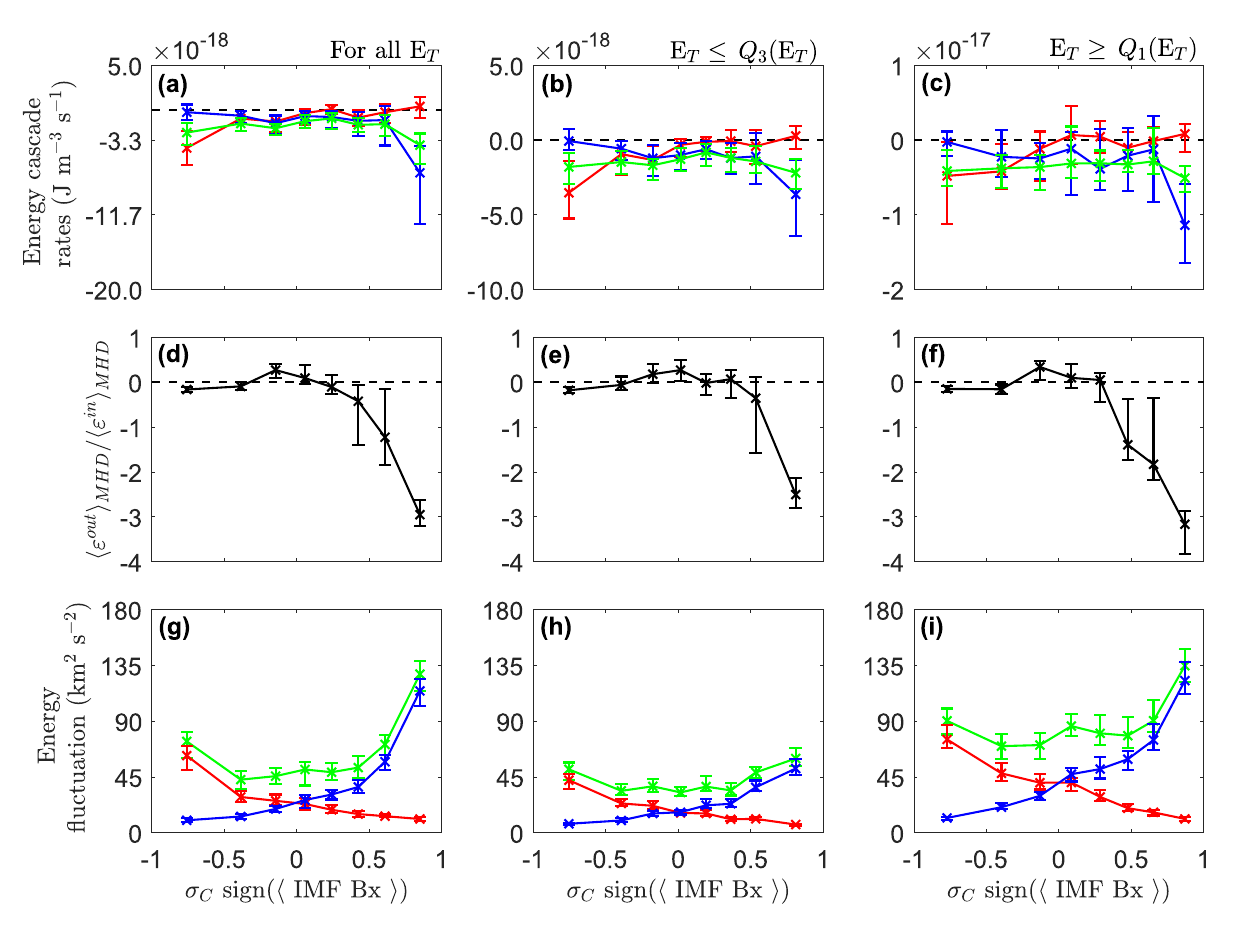}
\end{center}
\caption{(a-c) Median energy cascade rates at MHD scales as a function of $\sigma_c$ sign($\langle$IMF Bx$\rangle$) for all $E_{T}$, $E_{T}\leq Q_3(E_{T})\sim$ 120 km$^2\,$s$^{-2}$, and $E_{T}\geq Q_1(E_{T})\sim$ 30 km$^2\,$s$^{-2}$, respectively. Blue, red and green curves correspond to $\langle\varepsilon^{out}\rangle_{MHD}$, $\langle\varepsilon^{in}\rangle_{MHD}$ and $\langle\varepsilon^{T}\rangle_{MHD}$, respectively. (d-f) Median $\langle\varepsilon^{out}\rangle_{MHD}/\langle\varepsilon^{in}\rangle_{MHD}$ as a function of $\sigma_c$ sign($\langle$IMF Bx$\rangle$) for the same $E_{T}$ conditions shown in panels (a-c). (g-i) Median of the total energy and pseudo-energies of solar wind fluctuations as a function of $\sigma_c$ sign($\langle$IMF Bx$\rangle$) for the same $E_{T}$ conditions shown in panels (a-c). Blue, red and green curves correspond to $E_{out}$, $E_{in}$ and $E_{T}$, respectively. Vertical bars are associated with the $95\%$ confidence interval for the corresponding median.  }
\label{fig5}
\end{figure}

In addition, our observational results show a strong correlation between the energy cascade rate intensity at  MHD scales and the total energy of solar wind fluctuations, in agreement with results reported by \citet{H2017a}. 
The implied power-law dependence between $|\langle\varepsilon^{T}\rangle_{MHD}|$ and $E_{T}$ appears intuitive when $\langle\varepsilon^{T}\rangle_{MHD}$ is positive. Indeed, the more energy fluctuation energy available, the larger the energy that can be transferred to the dissipative scales of the system. The fact that we observe a similar dependence for cases with negative energy transfer rates is somewhat more difficult to interpret. The common explanation involves an inverse cascade rate of energy, however, what would the energy source be in that case? A non trivial plasma process could be present, working as a source of energy in these particular scales. Interestingly, \citet{H2017a} reported an analogous polynomial fit between the compressible cascade rate and the compressible energy component. This observational result motivates further studies on this matter. In particular, future numerical simulation studies would allow a parametric analysis of how each solar wind variable influences the energy cascade rate. These simulations should take into account the observed differences when analyzing solar wind turbulence upstream of Earth and Mars. 
For instance, it is interesting to note that the events with high total energy of fluctuations analyzed in this work do not appear evenly distributed as a function of the solar wind residual energy nor the cross-helicity (see, Figure \ref{fig4} (a)). Moreover, additional efforts should be made to determine the effects that the time interval size may have on the computation of the energy cascade rate at MHD scales, and how it impacts the sign \citep{St2009}. In this sense, it is important to note that we have used time intervals larger than at least one correlation timescales at Martian heliocentric distances \citep{A2020}. 

{ It is also worth mentioning that solar activity may influence the energy cascade rate \citep[e.g.,][]{St2009,Coburn_2012,Marino_2012}. Indeed, \citet{Marino_2012} reported that the average pseudo-energy transfer rate is correlated with solar activity, based on Ulysses, high latitude, observations of fast solar wind. In this regard, given that the analyzed MAVEN dataset covers roughly half of the solar cycle, solar activity could partly be responsible for variability in the computed energy transfer rate upstream of Mars bow shock. A comprehensive analysis of these potential effects requires observations spanning at least one complete solar cycle period and is left for a future study.}

Finally, solar wind turbulence studies around Mars would also benefit from continuous, high cadence, magnetic field and plasma observations in the region upstream of the Martian bow shock. Such data set would allow the determination of typical spatial scales (such us the correlation, Taylor scales and kinetic scales) and the energy cascade rate at different timescales. Furthermore, the continuous sampling of the pristine solar wind would allow one to analyze variability of these quantities with time intervals of different sizes, ranging from a few auto-correlation timescales (on the order of several tens of minutes) to days or even larger { and over different phases of the solar cycle.} Such observations could be provided by a future Heliophysics mission to the Lagrangian 1 point at Mars \citep{Lee2023C}.

\vspace{1cm}
The MAVEN project is supported by NASA through the Mars Exploration Program. N.R. is supported through a cooperative agreement with Center for Research and
Exploration in Space Sciences \& Technology II (CRESST II) between NASA Goddard Space Flight Center and University of Maryland College Park under award number
80GSFC21M0002.  MAVEN data are publicly available through the Planetary Data System (\url{https://pds-ppi.igpp.ucla.edu/index.jsp}).

\begin{table}
  \caption{Quartiles of distributions for $\sim$ 34 min and  $\sim$ 68 min intervals. These distributions consists of 1890 and 376 events, respectively.}
\[
\begin{array}{|c|c|c|c|c|c|c|}
\hline
\multirow{2}{*}{\text{Variable (unit)}} & \multicolumn{2}{c|}{\text{Lower quartile } (Q_1)}  & \multicolumn{2}{c|}{\text{Median }  (Q_2)} & \multicolumn{2}{c|}{\text{Upper quartile } (Q_3)} \\
\cline{2-7}
 & \sim34\, \text{min}& \sim68\, \text{min} & \sim34\, \text{min} & \sim68\, \text{min} & \sim34\, \text{min} & \sim68\, \text{min} \\
\hline
E_{in} \text{ (km}^2\text{ s}^{-2}\text{)} & 10 & 14 & 21 & 28 & 46 & 62 \\
E_{out} \text{ (km}^2\text{ s}^{-2}\text{)} & 12 & 18 & 28 & 38 & 68 & 90 \\
E_{T} \text{ (km}^2\text{ s}^{-2}\text{)} & 29 & 43 & 60 & 84 & 123 & 150 \\
\hline
\langle\text{IMF Bx}\rangle \text{ (nT)} & -0.93 & -0.84 & 0.03 & -0.04 & 1.19 & 1.01 \\
\sigma_r & -0.50 & -0.52 & -0.33 & -0.35 & -0.14 & -0.19 \\
\sigma_c \text{ sign}(\langle\text{IMF Bx}\rangle) & -0.25 & -0.22 & 0.15 & 0.16 & 0.52 & 0.50 \\
\hline
\langle \varepsilon^{in} \rangle_{MHD} \text{ (} \times 10^{-17} \text{ J m}^{-3} \text{ s}^{-1} \text{)} & -1.00 & -0.81 & -0.07 & -0.08 & 0.82 & 0.86 \\
\langle \varepsilon^{out} \rangle_{MHD} \text{ (} \times 10^{-17} \text{ J m}^{-3} \text{ s}^{-1} \text{)} & -1.53 & -0.98 & -0.11 & -0.12 & 0.83 & 0.66 \\
\langle \varepsilon^{T} \rangle_{MHD} \text{ (} \times 10^{-17} \text{ J m}^{-3} \text{ s}^{-1} \text{)} & -1.11 & -0.91 & -0.18 & -0.18 & 0.57 & 0.45 \\
\hline
\end{array}
\]
\end{table}

\end{document}